# File-based localization of numerical perturbations in data analysis pipelines

Ali Salari[1], Gregory Kiar[2,3], Lindsay Lewis[2], Alan C. Evans[2,3], Tristan Glatard[1]
[1]Concordia University, Montreal, Canada; [2]McGill University, Montreal, Canada; [3]Montreal Neurological Institute, Montreal, Canada.

*Abstract*—Data analysis pipelines are known to be impacted by computational conditions, presumably due to the creation and propagation of numerical errors. While this process could play a major role in the current reproducibility crisis, the precise causes of such instabilities and the path along which they propagate in pipelines are unclear. We present Spot, a tool to identify which processes in a pipeline create numerical differences when executed in different computational conditions. Spot leverages system-call interception through ReproZip to reconstruct and compare provenance graphs without pipeline instrumentation. By applying Spot to the structural pre-processing pipelines of the Human Connectome Project, we found that linear and non-linear registration are the cause of most numerical instabilities in these pipelines, which confirms previous findings.

## I. Introduction

Numerical perturbations resulting from variations in computational environments impact data analyses in various fields, but identifying the origin of these perturbations in complex pipelines remains challenging. In some cases, small perturbations resulting from changes in operating system versions [14], hardware [19], or parallelization parameters [9], result in substantially different analysis outcomes, due to the propagation and amplification of floating-point errors. While the existence of such numerical errors is well known [31], their impact on scientific computations has multiplied with the rise of the Big Data era, due to the sustained growth of data sets, the increasing complexity of analysis pipelines, and the diversification of computing infrastructures. To better understand and correct these effects, efficient tools are needed to assist pipeline developers in the comparison of results obtained across different conditions.

In neuroimaging, our primary application field, data analyses often consist of hundreds of computational processes – often coming from multiple toolboxes – that are aggregated to perform a specific function. For instance, the fMRIprep pipeline [10] assembles software blocks from FSL [18], AFNI [6], FreeSurfer [11] and ANTs [1] to provide a state-of-the art functional MRI processing tool with minimal user input. Another example are the pipelines of the Human Connectome Project [12] that combine tools from FSL and FreeSurfer to pre-process structural, functional and diffusion data from their uniquely high-fidelity open dataset. In both cases, pipelines leverage toolboxes that are widely trusted in the community, yet, at the same time substantial variations in results have been observed in these toolboxes resulting from minor data or infrastructure perturbations [15], [14], [22], [20], suggesting that further investigation of their numerical conditioning is required. For such complex pipelines, a lightweight solution has to be found to perform such evaluations with limited code instrumentation.

Numerical evaluations are traditionally performed using techniques such as interval arithmetics [16] that require complete code re-writes and are therefore barely applicable to complex pipelines. Recently, Monte-Carlo Arithmetic [26], [7] provided a practical way to evaluate the uncertainty of numerical results without the need to rewrite the application in a different paradigm. By perturbating floating-point computations, it introduces a controllable amount of noise in the pipelines, effectively sampling results from a random distribution. While this technique is very appealing, it suffers from two main issues that make it impractical at the scale of a complete pipeline. First, it requires that all software components be recompiled for MCA instrumentation, which is not always feasible. Second, it multiplies the execution time by a factor of 10 to 100, which is impractical when executions already take a few hours to complete.

We present Spot, a tool to identify the source of numerical differences in complex pipelines without instrumentation. Using system-call interception through the ReproZip tool [29], Spot traverses graphs of processes and intermediary files to pinpoint the pipeline components that are unstable across execution conditions. When differences start accumulating, effectively masking any further instability, it restores clean data copies through a set of wrapper scripts. Wrapper scripts are also used to restore temporary data that might have been deleted during the execution, and to disambiguate files that have been written by multiple processes. The remainder of this paper presents the design of Spot, and its application to pre-processing pipelines of the HCP project.

## II. Tool description

Spot identifies the components in a pipeline, at the resolution level of a system process, that produce different results in different execution conditions. First, a directed bipartite provenance graph is recorded for each pipeline execution, where nodes represent application processes and files, and edges represent read and write file accesses (Figure 1a). Second, transient files, i.e., files that are either deleted during pipeline execution or modified by multiple processes, are identified and disambiguated, resulting in a provenance DAG (Directed Acyclic Graph) in which file nodes have a single parent (in-degree of 1) (Figure 1b). DAGs produced in different conditions are then compared, in a step-by-step execution that prevents the propagation of differences in the pipeline (Figure 1c). The resulting labeled graph identifies the non-reproducible processes in the pipeline.

To ensure that a file can be unambiguously associated with the process that created it, we assume that the pipeline can be transformed such that:

1) Processes don't run concurrently;
2) Each process sequentially reads, computes, and writes.

In practice, pipeline processes may still run concurrently provided that they don't write concurrently to the same files. A process may also interleave file writes with computing, for instance when different file blocks are processed sequentially. However, only a single version of the file must eventually be made available to the other processes. In particular, in case a process deletes a file that it had created itself, this file must not be used by any other process. Finally, we also require that processes are associated to a command line (executable and arguments), to facilitate process instrumentation.

```bash
#!/usr/bin/env bash
if [ $# != 1 ]
then
    echo "usage: $0 <input_image.nii.gz>"
    exit 1
fi
# Parse argument, set output file names
input_image=$1
# Run FSL bet, put result in ${bet_output}
bet ${input_image} output.nii.gz
# Create binary mask
fslmaths output.nii.gz -bin output.nii.gz
echo "Voxels / volume in binarized brain mask:"
fslstats output.nii.gz -V > voxels.txt
# Remove temporary file
\rm output.nii.gz
```

Listing 1: Example pipeline that computes the volume of the brain from a T1 image.

### A. Recording provenance graphs

We use ReproZip [29] to capture: (1) the set of processes created by the pipeline, and (2) the set of files read and written by each process, including temporary files. ReproZip collects this information through the ptrace() system call, with no required instrumentation of the pipeline. Using the ReproZip trace, Spot reconstructs a provenance graph by creating process and file nodes and by adding directed edges corresponding to file reads and writes (Figure 1a). We assume that provenance graphs are identical for the ReproZip traces obtained from the same subjects in different operating systems.

Provenance graphs are often data-dependent, due to variations in input data that may trigger differing branching or looping patterns across executions, for example. Some of these differences can be neglected: for instance, when a data decompression step is present at the beginning of the execution for some subjects only. Other differences cannot: for instance, when entirely different processing paths are used for different datasets. Spot includes helpers to identify different instances of provenance graphs, such as supporting the clustering of process trees, where nodes are processes and edges are fork() or clone() system calls, using the tree edit distance [34] implemented in Python's zss package.

### B. Capturing transient files

We capture temporary files by replacing every process $P$ by a wrapper that first calls $P$ and then saves the produced temporary files to a read-only directory. This process replacement is done by pre-pending to the PATH environment variable a directory that contains a wrapper script named after the executable called by $P$.

Files written by multiple processes are disambiguated using a similar technique. For a file $F$ written by the processes in **P** = $\{P_1, \ldots, P_n\}$, we first check that processes in **P** do not write concurrently to $F$, which would violate our assumptions. Then, we replace every process $P_i$ by a PATH-based wrapper that first calls $P_i$ and then saves $F$ to a read-only directory. In this way, successive versions of $F$ are preserved for comparison. We finally update the provenance graph accordingly, so that all files in the graph have an in-degree of 1 (Figure 1b). This operation also makes the provenance graph acyclic, since we assumed that a process could only release a single version of a file.

### C. Labeling processes

After capturing transient files in the first condition (i.e. operating system, library version, etc.), we re-run the pipeline step by step in the second one to label processes. The output files created by a process in both conditions are compared: if no differences are found, the process is marked as reproducible; otherwise, the process is marked as non-reproducible, and the output files produced in the first condition are copied to the second one, to ensure that differences do not propagate further in the pipeline. Processes are instrumented transparently through a modification of the PATH variable similar to the one described previously. By default, differences in output files are identified by comparing file checksums. Other comparison functions can also be defined for specific file types, for instance to ignore file headers or file sections containing timestamps. Spot finally creates a labeled provenance graph highlighting non-reproducible processes.

Figure 1c illustrates a hypothetical incremental labeling of the example in Listing 1. Process bet2 is labeled as non-reproducible (red) as it produces files with differences. To prevent the propagation of these differences, the files produced by bet2 in Condition 2 are replaced with the files produced by bet2 in Condition 1. Processes fslmaths and fslstats are then executed and labeled as reproducible (green) as they produce files without differences.

The labeled graph can differ depending on the order of executions in which condition we capture transient files or execute the pipeline to pinpoint the propagation of differences. Therefore, we run the comparison in both condition orders, and we label a process as non-reproducible (red) if it creates different results in at least one condition order.

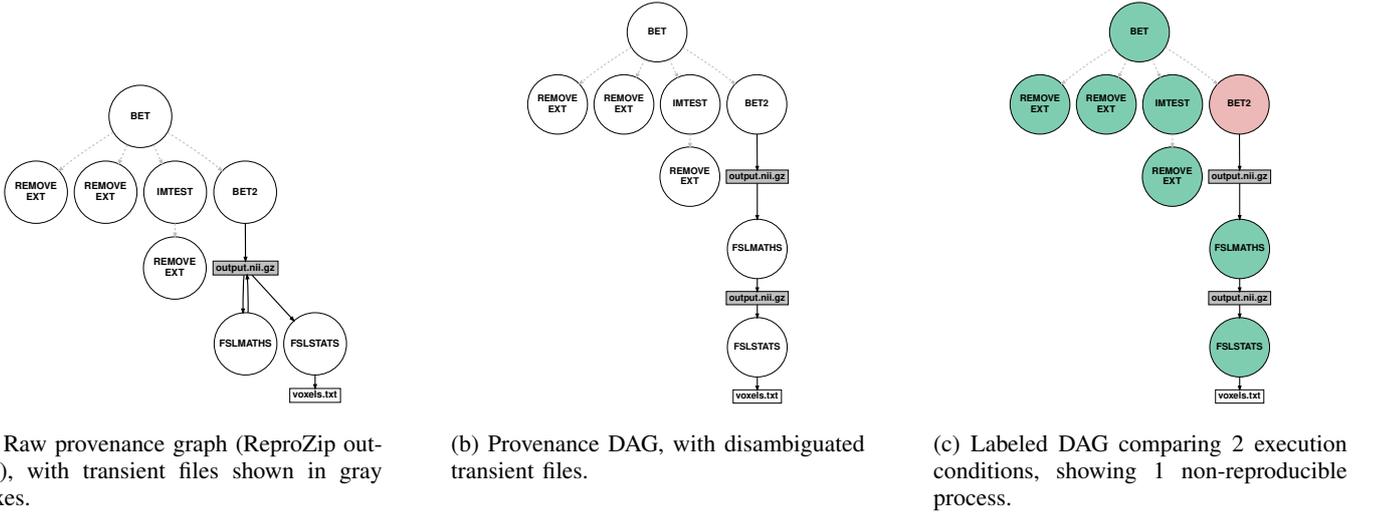

(a) Raw provenance graph (ReproZip output), with transient files shown in gray boxes.

(b) Provenance DAG, with disambiguated transient files.

(c) Labeled DAG comparing 2 execution conditions, showing 1 non-reproducible process.

Fig. 1: Provenance graphs created from the example pipeline in Listing 1. Processes are represented with circles, files with rectangles, and read/write accesses with plain edges. For convenience, the process tree is also shown, with gray dashed edges. Processes forked by `bet` were captured by ReproZip while they did not appear in Listing 1. Processed associated with executables located in `/usr/bin/` or `/bin/` are not shown.

*D. Implementation*

Spot is implemented in Python (>=3.6). In this work we used Spot version 0.2 and the following version of the Python package dependencies: NumPy v1.19.0 [25] and Pandas v1.0.5 [23], for data manipulations, SciPy v1.5.1 [33] and Scikit-learn v.0.23.1 [27] for the clustering of provenance graphs, Zss v1.2.0 [34] for tree distances, ReproZip v1.0.11 for the capture of provenance traces, Docker v17.05 [24] for the edition of container images, and Boutiques v0.5.25 [13] for uniform pipeline executions.

Software users will mostly have to interact with the Boutiques and ReproZip packages. Boutiques is a flexible description framework for containerized pipelines, required by the pipelines analyzed in Spot. It provides a JSON schema to describe inputs, outputs and their dependencies. Examples, tutorials and usage documentation are available at http://boutiques.github.io. ReproZip intercepts system calls to identify the files and processes involved in a pipeline execution. Before using Spot, users have to collect ReproZip traces of their pipeline executions. Examples in the Spot documentation include ReproZip provenance capture. More documentation on ReproZip is available at https://www.reprozip.org.

## III. EXPERIMENTS

We applied Spot to the minimal pre-processing pipelines released by the Human Connectome Project (HCP), a leading initiative in neuroimaging.

*A. HCP pipelines and dataset*

The HCP developed a set of pre-processing pipelines to process structural, functional, and diffusion MRI data acquired in the project. We focus on HCP pre-processing pipelines for structural data, and particularly on PreFreeSurfer and FreeSurfer. A detailed description of the analyses done by these pipelines is available in [12]. In summary, the PreFreeSurfer pipeline consists of the following steps:

- Gradient Distortion Correction (DC),
- Alignment and Anatomical Average (AAve), T1w(s), T2w(s),
- Anterior/Posterior Commissure Alignment (ACPC-A),
- Brain Extraction (BExt),
- Bias Field Correction (BFC),
- Atlas-Registration (AR).

And the FreeSurfer pipeline consists of the following:

- Image downsampling,
- T1w image registration,
- T1w image segmentation,
- Surface placement,
- Surface registration.

We randomly selected 20 unprocessed subjects from the HCP data release S500 available in the ConnectomDB repository as a subset of the 1200 Subject Release (see Supplementary Table S1). For each subject, available data consisted of 1 or 2 T1-weighted images and 1 or 2 T2-weighted images, with $256 \times 320 \times 320$ voxels of size $0.7 \times 0.7 \times 0.7$ mm. Acquisition protocols and parameters are detailed in [32].

*B. Data processing*

We built Docker images for the HCP pre-processing pipelines v3.19.0 (PreFreeSurfer and FreeSurfer) in CentOS 6.9 (Final) and CentOS 7.4 (Core), available on DockerHub. Container images contain the HCP software dependencies, including FSL (version 5.0.6), FreeSurfer (version 5.3.0-HCP, CentOS4 build), and Connectome Workbench (version 1.0).

We processed the 20 subjects with PreFreeSurfer and FreeSurfer, using the 2 CentOS versions. The PreFreesurfer results obtained in CentOS6 were used as the input of FreeSurfer in both conditions. We also used the ReproZip trace file captured in CentOS6 for labeling the processes in both pipelines. Each subject was processed twice on the same operating system to detect within-OS variability coming from pseudo-random operations. We compared pipeline results using FreeSurfer tools `mri_diff`, `mris_diff`, and `lta_diff`, to ignore execution-specific information such as file path or timestamps. To compare segmentations $X$ and $Y$, we used the Dice coefficient defined as follows:

$$DICE = \frac{2|X \cap Y|}{|X| + |Y|}$$

The Dice coefficient [8] is a commonly used metric to validate medical image segmentation. Dice values range from 0 to 1, with 1 indicating a perfect overlap between two segmentation results and 0 indicating no overlap. Alternatively, the Jaccard coefficient [17] could be used; there is a direct correspondence between both metrics.

## IV. Results

All experiments were run on a machine with a 3.4GHz, 8-core Intel Core i7 processor, 32GB of RAM, CentOS 7.3.1611, and Linux kernel version 3.10. The processing time, output file size, number of file accesses and number of processes observed in PreFreeSurfer and FreeSurfer are shown in Table I. The scripts and analyses used to create the figures in this section are available at https://github.com/big-data-lab-team/HCP-reproducibility-paper.

*1) Within-OS differences:* We did not observe any within-OS difference in PreFreeSurfer. In FreeSurfer, we identified 2 processes leading to within-OS differences due to the use of pseudo-random numbers: image registration with `mri_segreg`, and cortical surface curvature estimations with `mris_curvature`. Fixing the random seed used in FreeSurfer removed these differences.

*2) Between-OS differences in PreFreeSurfer:* We identified four types of subjects with different PreFreeSurfer provenance graphs (Table II). Differences between subject types came from different numbers of T1 and T2 images in the raw data. We verified that the provenance graphs were identical for all subjects of the same type, for both versions of CentOS.

Figure 2 shows the frequency of non-reproducible pipeline processes in PreFreeSurfer. The processes identified as non-reproducible were observed in linear registration with FSL `flirt` (in ACPC-Alignment, Brain Extraction, Distortion Correction, and Atlas Registration), in non-linear registration with FSL `fnirt` (in Brain Extraction and Atlas Registration), and in image warping with FSL `new_invwarp` (in Brain Extraction and Atlas Registration). Differences were also observed in image mean computations with FSL `maths` (in Anatomical Average). Figure 3 shows a complete PreFreeSurfer labeled DAG, localizing the observed differences in the entire pipeline, for a given subject.

Figure 4 compares `fnirt` results in Brain Extraction for a particular subject using the checkerboard pattern, a common method to illustrate the magnitude of the differences in registration results. Differences appear to be visually important, in particular in the areas framed in red, to the point that most experimenters would likely reject such a registration following visual quality control.

*3) Between-OS differences in FreeSurfer:* The only non-reproducible process identified by Spot in FreeSurfer was `mris_make_surfaces` (cortical and white matter surfaces generation), a dynamically-linked executable that produced different results for 10 out of 20 subjects.

However, FreeSurfer results still differ between conditions, due to the propagation of differences created in PreFreeSurfer. We observed the effect of this propagation in FreeSurfer results, as shown in Figure 5 for whole-brain segmentations. The Dice coefficients associated with the 44 regions segmented by FreeSurfer are shown in Figure 6, showing that Dice coefficients below 0.9 are observed in most regions, and particularly in the smallest ones. However, no significant correlation between the Dice values and the region sizes was found (Pearson's coefficient = 0.12, p-value = 0.43).

## V. Discussion

Our results provide insights on the reproducibility of neuroimaging pipelines, and on the relevance of the approach implemented in Spot for reproducibility studies.

### A. Key findings

Linear and non-linear registration with FSL were found to frequently lead to differences between results obtained with different operating systems. This does not come as a surprise given the instabilities associated with these processes. It also corroborates our previous findings in [14], where fMRI pre-processing with FSL was found to vary across operating systems starting from the motion correction step, a step that uses FSL's `flirt` tool internally. It would be relevant to investigate if the observed instability of registration processes generalizes to other toolkits, or if it remains specific to FSL. In view of the effect of small data perturbations in a variety of toolboxes and processes, such as cortical surface extraction using FreeSurfer and CIVET [22] or connectome estimation using Dipy [21], it is probable that this observation generalizes widely across toolboxes and requires a deeper investigation of the stability of linear and non-linear registration.

While only a handful or processes were found non-reproducible across the tested operating systems, the effect of such instabilities were found to propagate widely in the pipelines, and to substantially impact the segmentations created by FreeSurfer. This illustrates the need to conduct reproducibility studies on entire pipelines rather than isolated processes. It also highlights the need for a deeper stability analysis of pipeline processes.

As is shown in Figure 2, the reproducibility of a given tool may vary across subjects and across processing parameters. For instance, linear registration with `flirt` seems to be fully reproducible in the Anatomical Average sub-pipeline, while it is highly non-reproducible in ACPC Alignment. In Brain Extraction, the same tool was found reproducible for some subjects only. Therefore, reproducibility studies need to be performed on several subjects. While this is common practice to some extent in neuroimaging, software tests are often executed only on a single dataset to reduce the associated computational load. Our results show that pipeline tests should encompass enough subjects to cover execution paths adequately.

TABLE I: Execution statistics of the pipelines per subject.

|  | PreFreeSurfer | | FreeSurfer | |
|---|---|---|---|---|
|  | Mean | Standard error | Mean | Standard error |
| Processing time (mins) | 106.67 | 2.68 | 650.25 | 8.88 |
| Output file size (GB) | 2.8 | 0.10 | 4.15 | 0.15 |
| Number of file accesses | 94,089 | 2,645 | 62,729 | 984 |
| Number of processes | 8,731 | 198 | 4,031 | 47 |

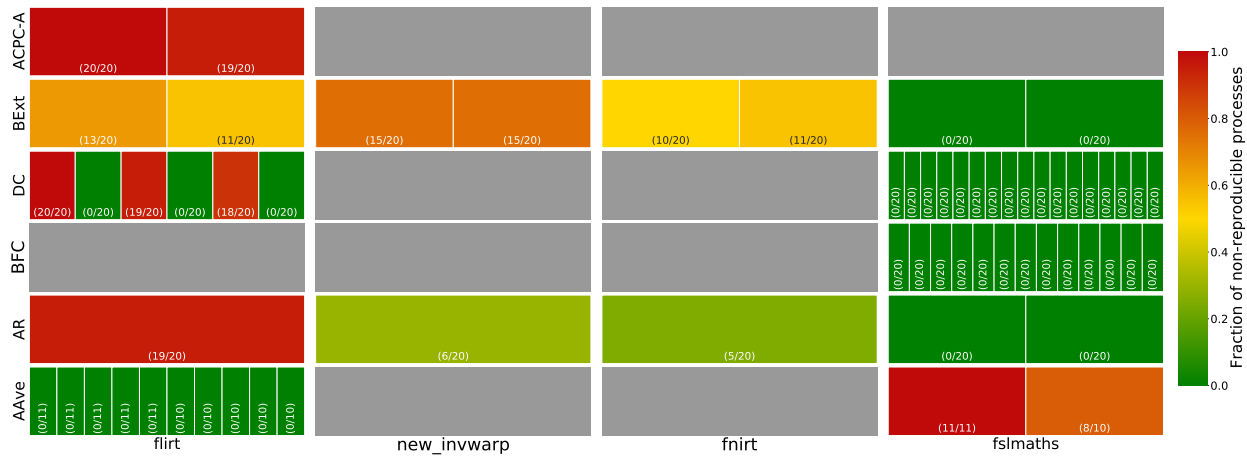

Fig. 2: Heatmap of non-reproducible processes across PreFreeSurfer pipeline steps. Each cell represents the occurrence of a particular command line in a pipeline step among Anatomical Average (AAve), Anterior/Posterior Commissure Alignment (ACPC-A), Brain Extraction (BExt), Bias Field Correction (BFC), or Atlas-Registration (AR). Cell labels indicate the fraction of subjects for which the corresponding process wasn't reproducible. For example, the `flirt` tool was invoked 6 times in step DC for each of the 20 subjects: 2 instances weren't reproducible in 19 subjects, 3 instances were always reproducible, and 1 instance wasn't reproducible in 17 subjects. Grey cells indicate that the process did not occur in the corresponding pipeline step.

Our results illustrate the type of variability that can be introduced in neuroimaging results due to operating system updates. The numerical noise introduced by operating system updates is realistic, as such updates are likely to occur throughout the time span of a neuroscience study, but it is also uncontrolled, as it originates in updates of low-level libraries by third-party developers. A possible method to study this problem more comprehensively would be to introduce controlled numerical perturbations in pipelines, which could be done by introducing noise either in the data, or in floating-point computations through Monte-Carlo Arithmetic [26]. The work in [21] discusses and compares these two techniques.

TABLE II: Types of provenance graphs in PreFreeSurfer.

| Type | Number of Subjects | Number of T1w images | Number of T2w images |
|---|---|---|---|
| 1 | 9 | 2 | 2 |
| 2 | 8 | 1 | 1 |
| 3 | 1 | 1 | 2 |
| 4 | 2 | 2 | 1 |

*B. Spot evaluation*

The processes identified by Spot as non-reproducible were all associated with dynamically-linked executables. This makes complete sense as statically-linked executables are not impacted by library updates. Moreover, the hypothetical effects of hardware or Linux kernel updates were not measured, as the different operating systems were deployed in Docker containers on the same host, that is, using the same kernel and hardware.

To evaluate the reproducibility of a pipeline, Spot needs to execute it 5 times in order to (1) record a first ReproZip trace, (2) save transient files in the first condition, (3) compare results in the second condition, and repeat steps (2) and (3) for the other order of execution. It might be possible to further reduce this overhead by executing at step (2) only the processes depending on transient files, and capturing the transient files for the second condition simultaneously at step (3).

The target users of the Spot tool are primarily pipeline developers and users who have technical skills for creating Docker containers and Boutiques JSON files. We demonstrated the applicability of our approach by evaluating two of the arguably most complex pipelines in neuroimaging. Technically, these pipelines consist of a mix of tools assembled from different toolboxes through a variety of scripts written in different languages. Our file-based approach, notably enabled by ReproZip, was able to analyze these pipelines without requiring their instrumentation, which saved a very substantial technical effort. The assumptions made on the pipeline structure, related to the absence of concurrent writes, were not violated in our analysis, and are likely to not impede Spot's applicability to the most common neuroimaging pipelines.

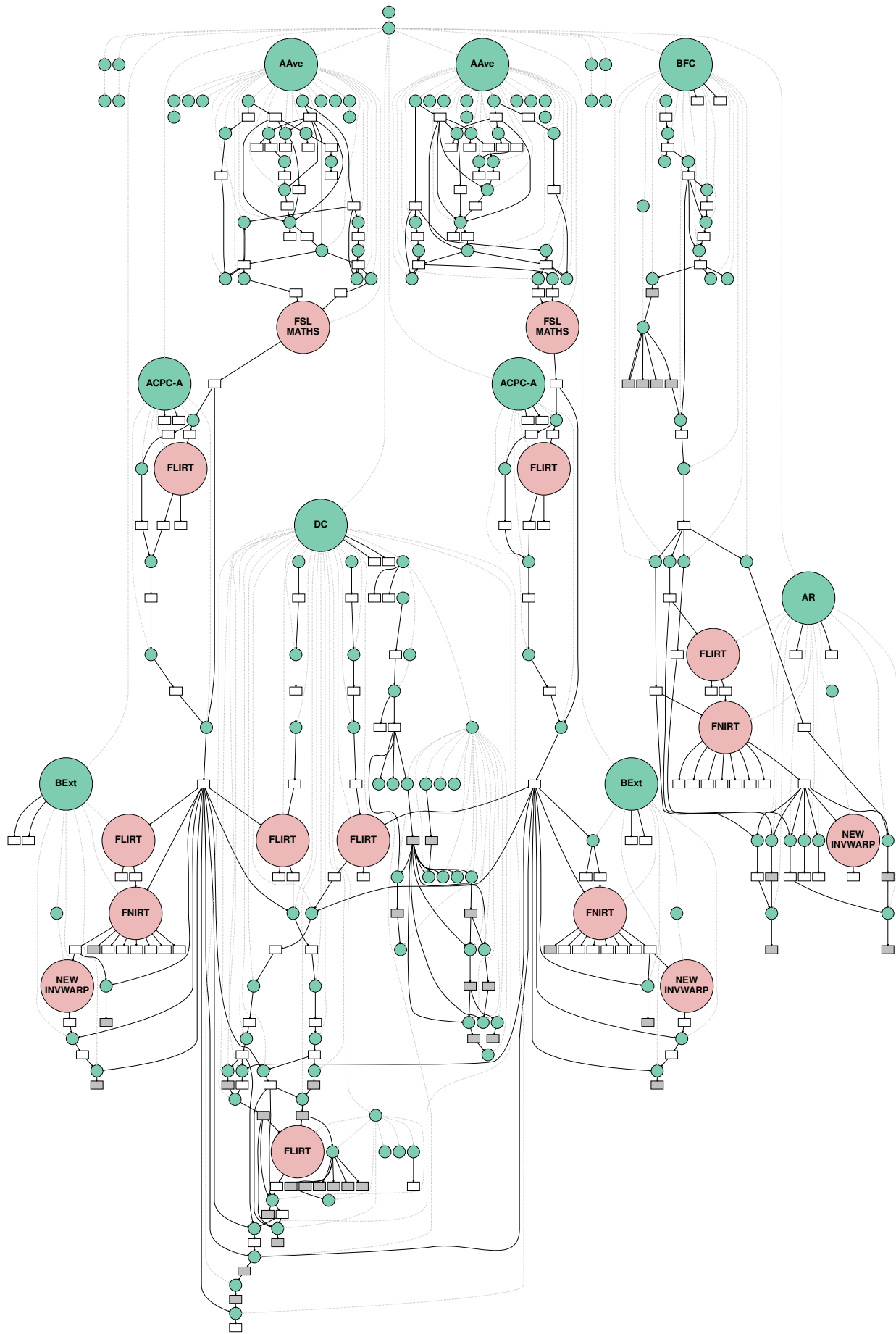

Fig. 3: A complete provenance graph from the PreFreesurfer pipeline. Node labels use the same abbreviations as in Figure 2. For better visualization, processes associated with commands in /bin or /usr/bin were omitted, as well as imtest, imcp, remove_ext, fslval, avscale, and fslhd.

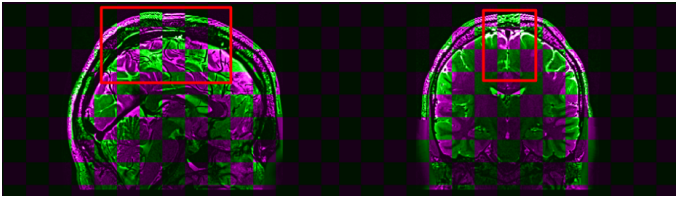

Fig. 4: Differences between T2 `fnirt` results in Pre-FreeSurfer's Brain Extraction (CentOS6 vs CentOS7). The colored squares indicate results obtained with CentOS6 (in purple) and CentOS7 (in green). The red boxes highlight regions with significant differences between the two OSes. An animated version of the comparison is available here for better visualization.

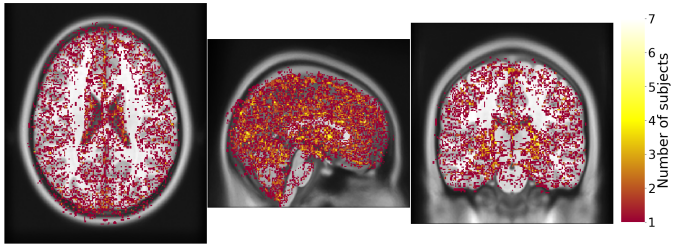

Fig. 5: Sum of binarized differences between whole-brain FreeSurfer segmentations obtained from PreFreeSurfer processings in CentOS6 vs CentOS7 (N=20). Segmentations were resampled and overlaid to the MNI152 volume template. Each voxel shows the number of subjects for which different results were observed between CentOS 6 and CentOS 7. An animated comparison of segmentations obtained for a particular subject is available here for better visualization.

Spot only tests pipeline reproducibility in the scope of a particular dataset. However, it is very plausible for pipeline processes to exhibit different reproducibility behaviors when executed on different datasets. Therefore, only the lack of reproducibility of a pipeline process could be guaranteed from an analysis with Spot, since proving reproducibility would require testing the pipeline on all possible datasets, in all possible environments, which is not feasible. Two elements could be considered in future work to address this issue. First, similar to conventional software testing, a code coverage metric could be developed to assess the fraction of the pipeline code involved in the tested dataset and parameters. This would quantify the representativity of the dataset and pipeline parameters used in the evaluation. Second, statistical risk models could be used to estimate the probability for a process to be reproducible, given a set of observations with no numerical differences. For instance, models described in [2] could be leveraged for this purpose.

File-based analyses also have limitations related to the granularity at which they operate. Indeed, differences can only be identified at the level of an entire operating-system process, which can correspond to arbitrary amounts of code. Narrowing down the analysis to particular libraries, functions, or even code sections would require another approach. Similarly, Spot would not be able to detect differences in data not saved in files but instead passed to subsequent processes in memory. A common scenario in neuroimaging pipelines is that tools return results in their standard output, which is parsed by the calling process and passed to subsequent ones through variables.

Computational environments are only one of many factors contributing to the on-going reproducibility crisis. In fact, sample size selection, publication bias, or methodological flexibility in the analysis are likely to have a stronger effect than numerical perturbations, although to our knowledge no evidence of this is available. We refer to the studies in [4], [5], [3], [20] for deeper analyses of the associated effects on neuroimaging analyses. It should also be noted that the effects of computational environments and these other factors manifest at different levels: referring to the terminology used in [28], computational environments are associated with reproducibility, the minimal standard by which identical results should be obtainable from identical data and parameters, while the other aforementioned factors belong to replicability, the ultimate standard by which independent experimenters should be able to draw similar conclusions from similar experiments. In practice, variability resulting from computational environments manifests during software testing (test results depend on execution platform), deployment on HPC systems (results obtained on local vs HPC systems differ), or software version updates (results obtained before vs after the update differ), while factors related to replicability impact the community more broadly. Ultimately, both reproducibility and replicability should be understood and improved.

## VI. CONCLUSION

We presented Spot, a tool to detect the source of numerical differences in complex pipelines executed in different computational conditions. Spot leverages system-call interception through the ReproZip tool, and therefore can be applied to the most complex pipelines without requiring their instrumentation. It is available at https://github.com/big-data-lab-team/spot under MIT license.

By applying Spot to the pre-processing pipelines of the Human Connectome Project, compared in different operating systems, we showed that between-OS differences are mostly originating in linear and non-linear image registration tools. Moreover, differences introduced during image registration propagate widely in the pipelines, leading to important variability in whole-brain segmentations.

Future work will investigate in more details the numerical stability of registration algorithms. Additionally, we plan on using Monte-Carlo arithmetic to inject controlled amounts of noise in pipelines and monitor uncertainty propagation and amplification in their results.

## VII. AVAILABILITY OF SOURCE CODE AND REQUIREMENTS

- Project name: Spot

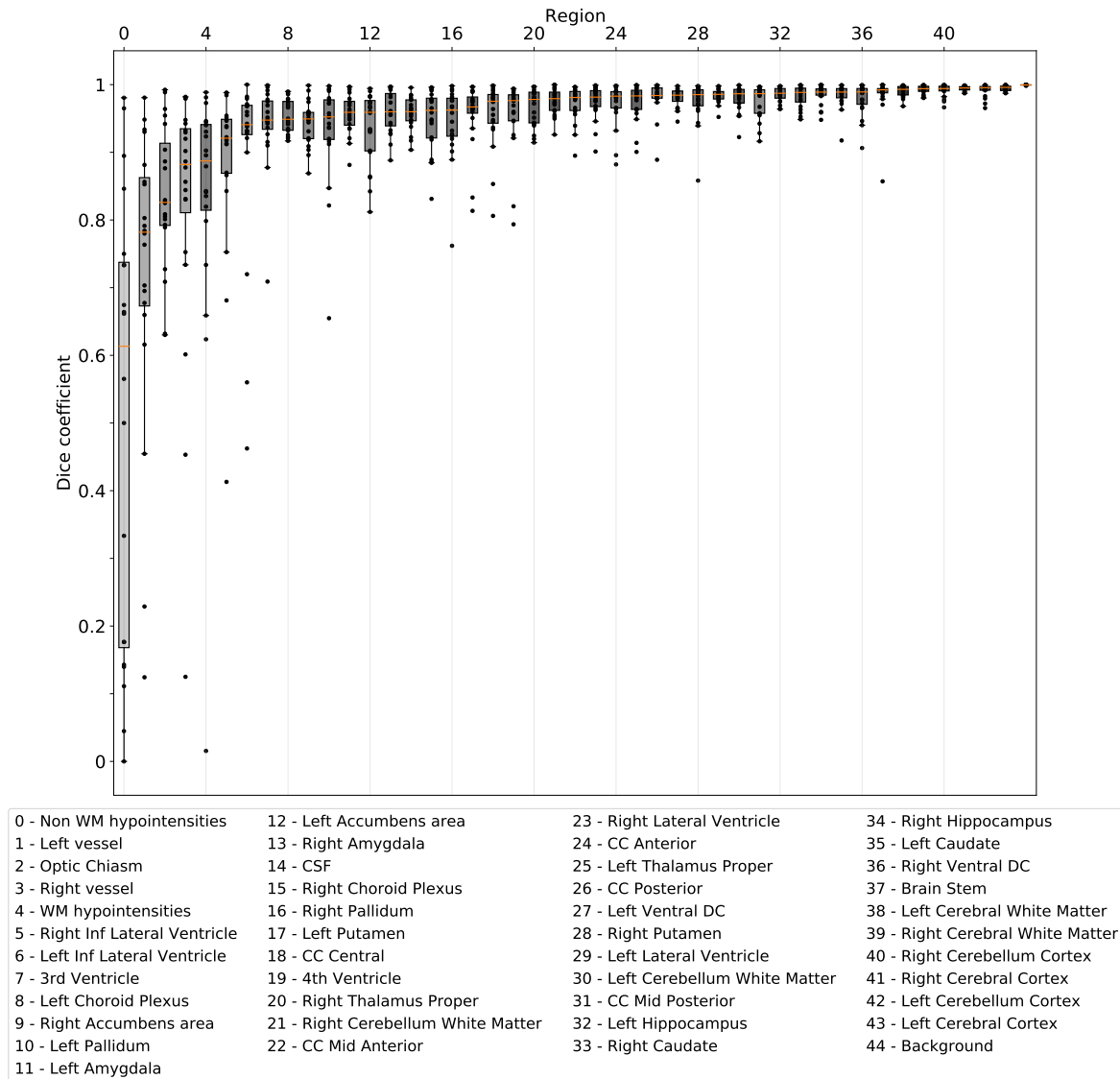

Fig. 6: Dice coefficients between regions segmented by FreeSurfer in CentOS6 vs CentOS7 (N=20), ordered by increasing median values. Each point represents the Dice coefficient between segmentations of a particular region obtained in CentOS 6 vs CentOS 7 for a given subject. Boxes brightness is proportional to the logarithm of the corresponding brain region size.

- Project home page: https://github.com/big-data-lab-team/spot
- Operating system: Linux
- Programming language: Python (3.6 or higher)
- Main dependencies: ReproZip, Docker, and Boutiques
- Other dependencies: see `setup.py`
- License: MIT License
- Biotools identifier: spottool
- SciCrunch ID: RRID:SCR_018915
- DOI: 10.5281/zenodo.3873219

## VIII. Availability of Supporting Data

Snapshots of our code and other supporting data are openly available in the GigaScience repository, GigaDB [30].

## IX. Supplementary Data

Supplementary materials include Supplementary Table S1 that contains a summary of the subjects used in the experiments.

## X. Acknowledgments

We thank the reviewers for their thorough and insightful review which greatly contributed to the improvement of the paper. We warmly thank Compute Canada (http://www.computecanada.ca) and Calcul Québec (http://www.calculquebec.ca) for providing the infrastructure used in our experiments. Data and pipelines were provided by the Human

4
Connectome Project, WU-Minn Consortium (Principal Investigators: David Van Essen and Kamil Ugurbil; 1U54MH091657) funded by the 16 NIH Institutes and Centers that support the NIH Blueprint for Neuroscience Research; and by the McDonnell Center for Systems Neuroscience at Washington University.

# File-based localization of numerical perturbations in data analysis pipelines

Ali Salari*, Gregory Kiar, Lindsay Lewis, Alan C. Evans, Tristan Glatard

## 1. SUPPLEMENTARY MATERIAL

We used unprocessed data from 20 subjects from the HCP data releases of the 1200 Subject project available in the ConnectomeDB repository. Table S1 shows a summary of the subjects.

**Table S1.** Summary of the subjects used in the experiments.

| Subject | Release[1] | Acquisition[2] | Gender | Age |
|---|---|---|---|---|
| 101309 | S500 | Q06 | M | 26-30 |
| 102008 | S500 | Q06 | M | 22-25 |
| 102311 | S500 | Q06 | F | 26-30 |
| 103414 | Q2 | Q02 | F | 22-25 |
| 103515 | Q1 | Q02 | F | 26-30 |
| 103818 | Q1 | Q01 | F | 31-35 |
| 105014 | S500 | Q05 | F | 26-30 |
| 105115 | Q2 | Q02 | M | 31-35 |
| 106319 | Q3 | Q03 | M | 26-30 |
| 106521 | S500 | Q06 | F | 26-30 |
| 107321 | S500 | Q04 | F | 22-25 |
| 107422 | S500 | Q07 | M | 22-25 |
| 108121 | S500 | Q04 | F | 26-30 |
| 108323 | S500 | Q04 | F | 26-30 |
| 113922 | S500 | Q04 | M | 31-35 |
| 140420 | Q2 | Q02 | F | 26-30 |
| 140824 | Q3 | Q04 | M | 31-35 |
| 140925 | S500 | Q04 | F | 22-25 |
| 142424 | Q3 | Q03 | M | 26-30 |
| 142828 | Q1 | Q01 | M | 31-35 |

[1] Release refers to the HCP data release in which this subject's data was initially published to ConnectomeDB
[2] Acquisition indicates the Quarter in which this subject's data was initially acquired